\documentclass[twocolumn,showpacs,preprintnumbers,amssymb]{revtex4}
\usepackage{graphicx, epsfig, epstopdf} 
\usepackage{amsmath, amsfonts, amssymb}
\usepackage{bm, array}
\usepackage[usenames]{color}
\usepackage[
      colorlinks=true,
      linkcolor=blue,
      urlcolor=blue,
      filecolor=blue,
      citecolor=red,
      pdfstartview=FitV,
      pdftitle={},
      pdfauthor={},
      pdfsubject={},
      pdfkeywords={},
      pdfpagemode=None,
      bookmarksopen=true
]{hyperref}

\def\be{\begin{equation}}
\def\ee{\end{equation}}
\def\beq{\begin{eqnarray}}
\def\eeq{\end{eqnarray}}

\begin{document}

\centerline{}
\title{Geodesic motion in equal angular momenta Myers--Perry--AdS spacetimes}

\author{T\'erence Delsate}
\email{terence.delsate@umons.ac.be}
\affiliation{Theoretical and Mathematical Physics Department,  University of Mons, UMONS, 20, Place du Parc, B-7000 Mons, Belgium}

\author{Jorge V. Rocha}
\email{jorge.v.rocha@tecnico.ulisboa.pt}
\affiliation{CENTRA, Departamento de F\'{\i}sica, Instituto Superior T\'ecnico, Universidade de Lisboa, Avenida Rovisco Pais 1, 1049 Lisboa, Portugal}

\author{Raphael Santarelli}
\email{santarelli@df.ufscar.br}
\affiliation{Departamento de F\'isica, Universidade Federal de S\~ao Carlos, P. O. Box 676, 13565-905, S\~ao Carlos, S\~ao Paulo, Brazil}

\date{\today}

\begin{abstract}
We study the geodesic motion of massive and massless test particles in the background of equally spinning Myers--Perry--anti-de Sitter (AdS) black holes in five dimensions. By adopting a coordinate system that makes manifest the cohomogeneity-1 property of these spacetimes, the equations of motion simplify considerably. This allows us to easily separate the radial motion from the angular part and to obtain solutions for angular trajectories in a compact closed form. 
For the radial motion we focus our attention on spherical orbits. In particular, we determine the timelike innermost stable circular orbits (ISCOs) for these asymptotically AdS spacetimes, as well as the location of null circular orbits.
We find that the ISCO dives below the ergosurface for black holes rotating close to extremality and merges with the event horizon exactly at extremality, in analogy with the four-dimensional Kerr case. For sufficiently massive black holes in AdS there exists a spin parameter range in which the background spacetime is stable against super-radiance and the ISCO lies inside the ergoregion. Our results for massless geodesics show that there are no stable circular null orbits outside the horizon, but there exist such orbits inside the horizon, as well as around overextremal spacetimes, i.e., naked singularities.
We also discuss how these orbits deform from the static to the rotating case.
 \end{abstract}


\maketitle

\section{Introduction}
\label{sec:Intro}

Geodesic structures convey useful information about the geometry of a given spacetime. In particular, these data give direct access to the main characteristics of black holes, such as its shadow~\cite{Chandrasekhar:1985kt}, scattering and absorption cross sections~\cite{Andersson:2000tf} or its stability properties~\cite{Berti:2009kk}.
On the other hand, geodesic motion is a first approximation to the two-body problem in general relativity, in which the particle following the geodesic is assumed not to backreact on the spacetime, or does so in a negligible way. This is precisely the scheme used for modelling extreme mass-ratio inspirals of binary systems~\cite{AmaroSeoane:2007aw}.

In this paper, we investigate the geodesic structure of five-dimensional equal angular momenta Myers--Perry black holes  with a negative cosmological constant. Previous higher-dimensional studies involved either asymptotically flat spacetimes (possibly with rotation) or spacetimes with a nonvanishing cosmological constant but static. This is, to our knowledge, the first work considering geodesics of higher-dimensional anti-de Sitter (AdS) space involving rotation. It can be regarded as a generalization of the study~\cite{Kagramanova:2012hw} which considered a vanishing cosmological constant, and which we are able to obtain as a limit. The present work may also be viewed as an extension of Ref.~\cite{Hackmann:2010zz} to five dimensions and of Ref.~\cite{Hackmann:2008tu} to include rotation.

The problem we tackle is known to be integrable. This was proved by Page {\it et al.} in Ref.~\cite{Page:2006ka}, and the separability of the Hamilton--Jacobi equation was demonstrated in Ref.~\cite{Frolov:2006pe}. Thus, the equations of motion can be reduced to first-order differential equations. However, these works say nothing about the geodesics themselves.

Let us briefly review the relevant results on black hole geodesics from earlier literature. 
Regarding nonrotating black holes it is well known~\cite{Chandrasekhar:1985kt} that the (four-dimensional) Schwarzschild geometry possesses stable timelike circular orbits for $r>6M$, unstable timelike circular orbits for $3M<r<6M$, and an unstable null circular orbit at $r=3M$. Here $r$ is the Schwarzschild radial coordinate and $M$ is the mass of the black hole.
Moving to spacetime dimensions $d\geq5$, all timelike or null circular geodesics outside the black hole horizon become unstable~\cite{Tangherlini:1963bw}. For Schwarzcshild--AdS$_d$ there can exist stable timelike circular orbits, but null circular orbits are still unstable~\cite{Cardoso:2008bp}.
Turning our attention to rotating backgrounds, it has been shown that all circular (either timelike or null) orbits on the equatorial plane of Myers--Perry are unstable in $d=5$ dimensions~\cite{Frolov:2003en}. This result generalizes to $d\geq5$ if one restricts to singly spinning geometries~\cite{Cardoso:2008bp}. Geodesics in asymptotically flat $d=5$ Myers--Perry (possibly with two independent spin parameters) have also been studied recently by Diemer {\it et al.}~\cite{Diemer:2014lba}, which confirmed the absence of stable bound orbits outside the event horizon. However, unstable circular orbits were identified both inside and outside the event horizon, and there were even found stable bound orbits within the Cauchy horizon.
Finally, singly spinning Myers--Perry black holes in dimensions greater than 5 can have arbitrarily large angular momentum~\cite{Emparan:2008eg}. In relation to this, Ref.~\cite{Igata:2014xca} pointed out that stable bound orbits exist outside the event horizon of such ultraspinning spacetimes in six dimensions. Moreover, it was argued that this feature should extend to all higher even-dimensional Myers--Perry geometries, rotating in a single plane and with a sufficiently large spin.

Besides being an interesting problem in itself, a strong motivation to study geodesics in anti-de Sitter is provided by its connection with conformal field theories in the context of the AdS/CFT correspondence~\cite{Maldacena:1997re, Witten:1998qj}. In this vein, the consideration of spacelike geodesics acquires particular relevance since it is intimately connected with the computation of $n$-point functions in the boundary field theory~\cite{Kraus:2002iv}. However, null geodesics also play an important role in the reconstruction of bulk spacetime (or a subregion thereof) from boundary data~\cite{Bousso:2012mh}.
Another general motivation to consider geodesics around black holes is that it gives a first glance at the properties of dynamical processes that have the black hole spacetime in question as an equilibrium configuration, using analytic or simple semianalytic techniques. Indeed, relations between the stability of circular null geodesics and the quasinormal mode spectrum of asymptotically flat black holes are known to exist (see \cite{Cardoso:2008bp} and references therein). It is, however, not clear how this connection extends to the case of asymptotically AdS spacetimes.

Here we take advantage of the symmetry enhancement due to the fact that the angular momenta are equal to easily separate the geodesic equations. The five-dimensional Myers--Perry black hole has generically a $\mathbb{R}\times U(1)^2$ isometry group, where the $U(1)$'s are associated with the two independent planes of rotation. When the angular momenta are equal, the spatial isometry gets enhanced to $U(2)\sim U(1)\times SU(2)$~\cite{Frolov:2002xf} (see also Ref.~\cite{Kunduri:2006qa}). This simplifying property has been used in several works in different contexts~\cite{Murata:2007gv,Murata:2008yx,Dias:2010eu,Dias:2010ma,Dias:2011at,Delsate:2014iia,Jorge:2014kra}. It follows that we find five conserved charges, associated with the generators of the symmetry (and the time translation), allowing one to completely separate the geodesic equations.
Our approach is different---but equivalent---to that of~\cite{Kagramanova:2012hw}, which is based on the Hamilton--Jacobi equations for the geodesics. However, given the higher degree of manifest symmetry in the coordinate system chosen, our equations of motion become somewhat simpler.

This paper is organized as follows. In Sec.~\ref{sec:background}, we review the geometry of Myers--Perry black holes in odd spacetime dimensions with equal angular momenta. In Sec.~\ref{sec:Static}, we discuss the symmetries of the line element in the nonrotating case. In particular, we identify the isometry subgroup that is left unbroken by the rotation and discuss the angular motion in detail in Sec.~\ref{sec:angular}. In Sec.~\ref{sec:ISCO}, we present the timelike innermost stable circular orbit. In Sec.~\ref{sec:rotating}, we turn on rotation and perform the same analysis for the case of timelike and null geodesics. We summarize and present our conclusions in Sec.~\ref{sec:Conc}. The derivation and separation of the geodesic equations using a Hamilton--Jacobi treatment is included in the Appendix~\ref{sec:HJ}.

\bigskip
\section{Background geometry: equally spinning Myers--Perry--AdS spacetime}
\label{sec:background}

We take as our background spacetime an equal angular momenta Myers--Perry solution, possibly with a cosmological constant, in $d=2N+3$ dimensions, with $N\geq1$ an integer~\cite{Myers:1986un,Gibbons:2004js}. This family of geometries has enhanced symmetry and their line element depends essentially only on a radial coordinate~\cite{Kunduri:2006qa}:
\beq
  ds^2 &=& g_{\mu\nu} dy^\mu dy^\nu = - f(r)^2 dt^2 + g(r)^2 dr^2 + r^2 \widehat{g}_{ab} dx^a dx^b \nonumber\\
&& + h(r)^2 \left[ d\psi + A_a dx^a - \Omega(r) dt \right]^2,
\label{eq:metric}
\eeq
where
\begin{flalign}
&  g(r)^2 = \left( 1 + \frac{r^2}{\ell^2} - \frac{2M}{r^{2N}} + \frac{2M a^2}{r^{2N}\ell^2} + \frac{2M a^2}{r^{2N+2}} \right)^{-1}\,, 
\label{eq:metricfuncs1}\\
&  h(r)^2=r^2\left(1+\frac{2M a^2}{r^{2N+2}} \right)\,, \quad \ \Omega(r)=\frac{2M a}{r^{2N} h(r)^2}\,,
\label{eq:metricfuncs2}\\
& f(r)=\frac{r}{g(r) h(r)}\,.
\label{eq:metricfuncs3}
\end{flalign}
%
The $x^a$ are $2N$ coordinates on the complex projective space $CP^N$, $\widehat{g}_{ab}dx^a dx^b$ is the Fubini--Study metric on such a manifold and $A=A_a dx^a$ denotes the associated K\"ahler potential. Expressions for $\widehat{g}_{ab}$ and $A_a$ can be obtained iteratively in $N$~\cite{Hoxha:2000jf}. The coordinates $y^\mu$ cover the whole spacetime and run over $\{t,r,\psi,x^a\}$. The angular coordinate $\psi$ parametrizes the $S^1$ fiber and has periodicity $2\pi$.

The metrics~\eqref{eq:metric} are solutions of the vacuum Einstein equations with a cosmological constant, proportional to
a length scale $-\ell^{-2}$,
\be
  R_{\mu\nu} = -2(N+1)\ell^{-2} g_{\mu\nu}\,,
\ee
and the limit $\ell\to\infty$ reproduces the asymptotically flat case.
Here and throughout the whole article we are setting Newton's constant $G=1$.

Finally, $M$ and $a$ stand for the mass and spin parameters, respectively. We take $a$ to be non-negative, without loss of generality.

The event horizon occurs at the largest real root of $g^{-2}$ (if it exists, this will depend on $M$ and $a$), and its spatial sections have the geometry of a homogeneously squashed $(2N+1)$-sphere. If $g^{-2}$ possesses no real positive roots then the metric~\eqref{eq:metric} describes a naked singularity.
%

The solutions as presented above apply to all odd spacetime dimensions $d\geq5$ but from now on we will restrict to the five-dimensional case, corresponding to $N=1$. Henceforth, we will denote the $x^a$ coordinates parametrizing $CP^1$ by $x^1\equiv\theta$ and $x^2\equiv\phi$. In this case, the Fubiniy--Study metric and the K\"ahler potential appearing in~\eqref{eq:metric} are given by
\beq
&& \widehat{g}_{ab}dx^a dx^b=\frac{1}{4}\left(d\theta^2+\sin^2\theta\, d\phi^2\right)\,, \\
&& A=A_a dx^a= \frac{1}{2}\cos\theta\, d\phi\,.
\eeq

\section{Geodesics of the nonrotating Schwarzschild--AdS black hole}
\label{sec:Static}

Before investigating the geodesic structure of the equal angular momenta Myers--Perry black hole, let us analyze the nonrotating case. These results are not new, since this is nothing but the (five-dimensional) Schwarzschild--AdS black hole, expressed in coordinates highlighting the angular isometry. Indeed, geodesics of higher-dimensional Schwarzschild--AdS spacetimes were studied in Ref.~\cite{Hackmann:2008tu}.

The nonrotating case is characterized by $h(r) = r,\ \Omega(r) = 0$ and $a=0$ in all the nontrivial metric functions. 
The isometry of the angular part is enhanced, compared to the rotating case, to $SU(2)\times SU(2)$. The Killing vectors of the line element~\eqref{eq:metric} are the following:
\begin{eqnarray}
&& \hspace{-0.3cm} \xi_t = \partial_t\,, \nonumber\\
&& \hspace{-0.3cm} \xi_1 = 2\left[ -\cos(2\psi) \partial_\theta  +\sin (2\psi)\left(\frac{1}{2} \cot\theta\, \partial_\psi -\csc\theta\, \partial_\phi \right) \right] , \nonumber\\
&& \hspace{-0.3cm} \xi_2 = 2\left[ \sin(2\psi) \partial_\theta  +\cos (2\psi)\left( \frac{1}{2} \cot\theta\, \partial_\psi -\csc\theta\, \partial_\phi \right) \right] , \nonumber\\
&& \hspace{-0.3cm} \xi_3 = \partial_\psi\,, \nonumber\\
&& \hspace{-0.3cm} \chi_1 = 2\left[ \cos\phi\, \partial_\theta -\sin\phi \left(\cot\theta\, \partial_\phi -\frac{1}{2} \csc\theta\, \partial_\psi\right) \right] , \nonumber\\
&& \hspace{-0.3cm} \chi_2 = 2\left[ - \sin\phi\, \partial_\theta -\cos\phi \left(\cot\theta\, \partial_\phi -\frac{1}{2} \csc\theta\, \partial_\psi\right) \right] , \nonumber\\
&& \hspace{-0.3cm} \chi_3 = 2\partial_\phi\,.
\label{eq:killing}
\end{eqnarray}
The $\chi_i$ and $\xi_i$ are the generators of two copies of $SU(2)$ and encode the isometry group of the angular sector\footnote{With the normalization adopted, the generators satisfy the standard $SU(2)$ commutation relations, namely $[\chi_i,\chi_j]=2\varepsilon^{ijk}\chi_k$ and $[\xi_i,\xi_j]=2\varepsilon^{ink}\xi_k$.}. Note that under the transformation $\phi\longleftrightarrow 2\psi+\pi$ the generators $\chi_i$ and $\xi_i$ get interchanged, with $i=1,2,3$.

The above seven Killing vectors lead to seven conserved charges. Clearly, only a subset will be independent, and these will include quantities representing the energy and the two independent angular momenta in each plane of rotation of the spacetime. 
The Killing equation, together with the geodesic equation for the velocity $u$, implies that if $\zeta$ is a Killing vector field then $u_\mu \zeta^\mu$ is constant over time. We thus define
\be
E = -u_\mu\xi_t^\mu\,, \quad G_i =u_\mu\xi_i^\mu\,, \quad L_i = u_\mu\chi_i^\mu\,,
\ee
where $u^\mu = \dot x^\mu$ is the velocity of the geodesic and the overdot denotes a derivative with respect to the proper time.
This leads to
\begin{eqnarray}
&&\dot t = \frac{E}{f(r)^2}\,, \nonumber\\
&&\dot\psi = \frac{\csc\theta}{r^2} (L_1 \sin\phi+L_2 \cos\phi)\,, \nonumber\\
&&\dot \theta =\frac{2}{r^2} (L_1\cos\phi-L_2 \sin\phi)\,, \nonumber\\
&&\dot \phi = \frac{2 }{r^2}\big(L_3 - \cot\theta(L_1\sin\phi + L_2 \cos\phi)\big)\,,
\label{eq:geoU}
\end{eqnarray}
with the additional constraints
\be
\vec G = -R_z(2\psi+\pi)R_x(\theta+\pi) R_z(\pi-\phi) \vec L\,,
\label{eq:consa0}
\ee
where $R_z$ and $R_x$ are the rotation matrices around $z$ and $x$, respectively:
\beq
R_x(\alpha) &=& \left(\begin{array}{ccc}
            1 & 0 & 0 \\
            0 & \cos \alpha & \sin \alpha \\
            0 &  -\sin \alpha& \cos \alpha \\
            \end{array}\right),\ 
\nonumber\\
R_z(\alpha) &=& \left(\begin{array}{ccc}
             \cos \alpha & \sin \alpha & 0\\
             -\sin \alpha& \cos \alpha & 0\\
             0 & 0 & 1
            \end{array}\right).
\eeq
It is easy to show that relation \eqref{eq:consa0} holds when the geodesic equations are satisfied. 
Most importantly, this relation implies $\vec G^2 = \vec L^2$. In other words, the Casimir invariant of the two $SU(2)$ algebras must be equal.

The radial motion is obtained by imposing $u^2 = -\kappa$, where $\kappa = 0$ ($\kappa=1$) for null (timelike) geodesics, leading to an equation of the form
\be
\dot r^2 + V(r) = 0\,,
\ee
where 
\be
V(r) = \frac{1}{g(r)^2}\left(\frac{\vec L^2}{r^2} +\kappa -\frac{E^2}{f(r)^2}\right) .
\label{eq:geoda0}
\ee

\subsection{Motion in the angular sector}
\label{sec:angular}

In this section, we derive an exact solution for the geodesic orbits in the $(\theta,\phi)$ plane. We also analyze the motion in the $(\psi,\phi)$ plane, for which an analytic solution cannot be found, except in particular cases.

Using the geodesic equations~\eqref{eq:geoU}, we find
\be
\frac{d\theta}{d\phi} = \frac{\dot\theta}{\dot\phi} = \frac{L_1\cos\phi-L_2\sin\phi}{L_3-\cot\theta (L_1\sin\phi+L_2\cos\phi)}\,.
\label{eq:ThetaPhiOrbit}
\ee
Defining
\be
\zeta(\phi) \equiv L_1\sin\phi + L_2\cos\phi\,,
\ee
we can recast Eq.~\eqref{eq:ThetaPhiOrbit} in the following form,
\be
\frac{d(\cos\theta)}{d\phi} = \frac{-\zeta'(\phi) \sin^2\theta}{L_3\sin\theta - \zeta(\phi)\cos\theta}\,,
\ee
which, upon integration, yields the general solution
\be
\cos\theta = \frac{c_1 L_3 \pm \sqrt{\zeta(\phi)^2 \big(\zeta(\phi)^2+L_3^2-c_1^2\big)}}{\zeta(\phi)^2+L_3^2}\,,
\label{eq:soltheta_c1}
\ee
where $c_1$ is a constant of integration. In fact, this constant of integration is none other than the constant of motion $G_3$. This follows directly from the third component of the constraints~\eqref{eq:consa0}, which yields \footnote{The third component of the vector constraints~\eqref{eq:consa0} is the only one that survives in the rotating case, and indeed it exactly reproduces what we will find in Sec.~\ref{sec:rotating}, Eq.~\eqref{eq:constraintG3}.}
\be
\zeta(\phi)=G_3\csc\theta- L_3 \cot\theta\,.
\ee
By squaring this relation and then solving for $\cos\theta$ we immediately obtain
\be
\cos\theta = \frac{G_3 L_3 \pm \sqrt{\zeta(\phi)^2 \big(\zeta(\phi)^2+L_3^2-G_3^2\big)}}{\zeta(\phi)^2+L_3^2}\,.
\label{eq:soltheta}
\ee
The reality of Eq.~\eqref{eq:soltheta} is guaranteed if $|G_3|\leq |L_3|$ \footnote{For orbits that complete a full revolution around the $\phi$ direction there always exist two values for this angular coordinate at which the function $\zeta$ vanishes. More specifically this occurs when $\tan\phi=-L_2/L_1$.}. However, this condition is not always satisfied: the constraints~\eqref{eq:consa0} only impose that $|G_3|\leq |\vec G| = |\vec L|$. Conversely, whenever $|G_3|>|L_3|$ the orbits cannot span the whole domain of the $\phi$ coordinate and instead must have turning points (in the angular motion) at $\phi=\phi_*$ such that $\zeta(\phi_*)^2=G_3^2-L_3^2$ \footnote{There is also the possibility that the orbit asymptotes to $\phi\to\phi_*$ as $t\to\pm\infty$ and $r\to\pm\infty$.}.

We may evoke the spherical symmetry of the spacetime (in this section only) to orient our coordinate system in such a way that the orbits go through the north pole of the 2-sphere parametrized by $(\theta,\phi)$ when $\zeta(\phi)=0$. This choice then imposes $G_3=L_3$. In this case, solution~\eqref{eq:soltheta} with the choice of the $+$ sign trivializes to $\cos\theta=1$. For the other branch (with the $-$ sign), the choice $G_3=L_3$ effectively determines an initial condition $\theta(\phi=0)$, and the solution~\eqref{eq:soltheta} reduces to a simple expression,
\be
\cos\theta = \frac{L_3^2 - (L_1 \sin\phi +L_2 \cos\phi )^2}{L_3^2+(L_1 \sin\phi +L_2 \cos\phi)^2}\,.
\label{eq:solthetasim}
\ee
As we will see in the next section, the geodesic motion in the $(\theta,\phi)$ plane is unaffected by the rotation, and therefore the general solution~\eqref{eq:soltheta} remains the same. The equation for $\psi$ will be, however, modified. 

\bigskip

In general, an analytic solution cannot be obtained for $\psi(\phi)$, not even when restricting to the nonrotating case.
We may nevertheless follow the procedure above to determine an expression for $\psi'(\phi)$ from the geodesic equations~\eqref{eq:geoU},
\beq
\frac{d\psi}{d\phi} = \frac{\dot\psi}{\dot\phi} &=& \frac{1}{2\sin\theta} \, \frac{L_1\sin\phi+L_2 \cos\phi}{L_3 - \cot\theta (L_1 \sin\phi+L_2 \cos\phi)} \nonumber\\
&=& \frac{\zeta(\phi)}{2\left(L_3\sin\theta - \zeta(\phi)\cos\theta\right)}\,,
\label{eq:PsiPhiOrbit}
\eeq
which, upon replacing the solution for $\theta(\phi)$, yields an ordinary differential equation for $\psi(\phi)$. For the case $G_3=L_3$ we get a very simple solution [depending on the choice of sign in Eq.~\eqref{eq:soltheta}]:
\be
\psi(\phi)=\mp\frac{\phi}{2}+c_2\,,
\ee
where $c_2$ is a constant of integration. We may impose the initial condition $\psi(\phi=0)=0$, without loss of generality. For generic choices of $G_3$ the solution for $\psi(\phi)$ can be found by a simple numerical integration of~\eqref{eq:PsiPhiOrbit}.

We illustrate the orbits of the nonrotating five-dimensional Schwarzschild background (projected onto the angular sector) in Figs.~\ref{fig:thetapsiphi1}--\ref{fig:thetapsiphi3}, where we plot the solution~\eqref{eq:soltheta} and the result of the numerical integration of~\eqref{eq:PsiPhiOrbit} for different combinations of $G_3$, $L_1$ and $L_2$, while fixing $L_3=1$. Figure~\ref{fig:thetapsiphi2} exemplifies the case discussed above, in which the coordinate $\phi$ possesses two turning points.

\begin{widetext}

\begin{figure}[h]
\centering
 \includegraphics[width=0.75\textwidth]{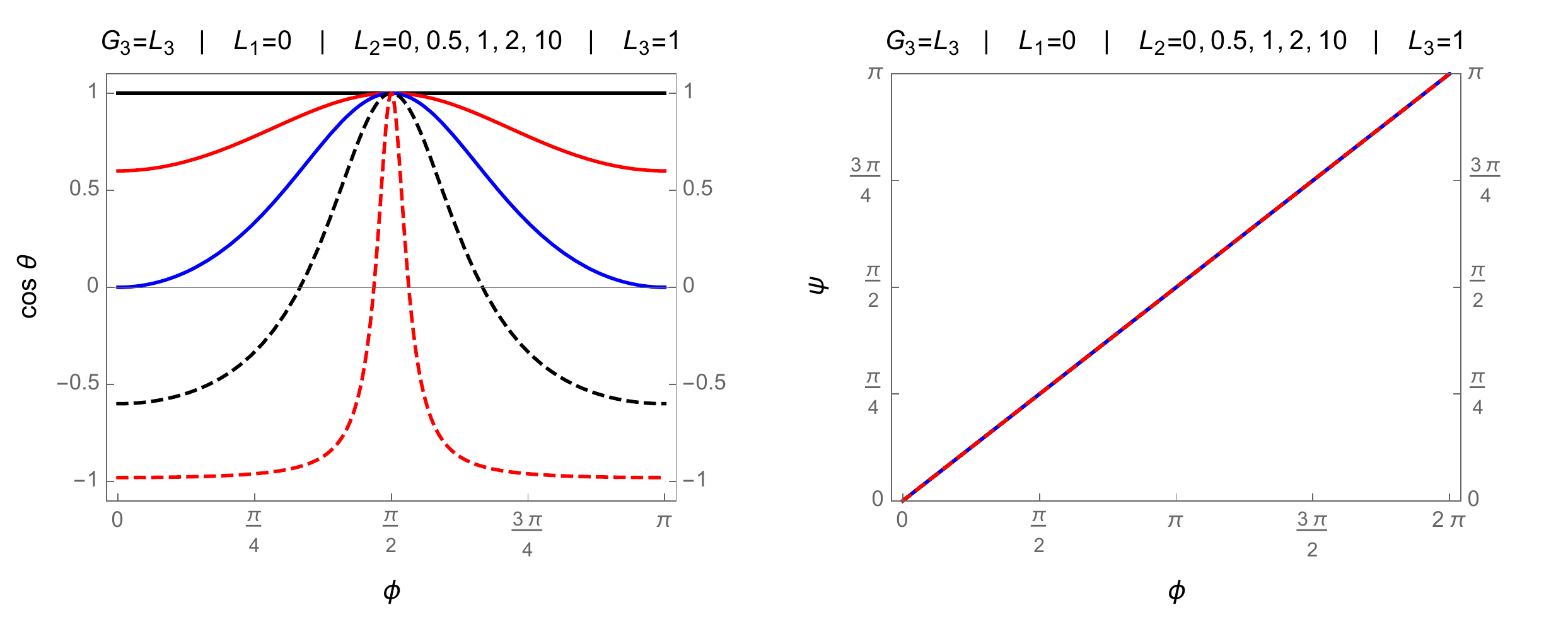}
 \vspace{-0.6cm}
 \caption{Left panel: Geodesic motion in the $(\theta,\phi)$ plane for $G_3=L_3,\ L_3=1,\ L_1=0$ and $L_2 = 0,0.5,1,2,10$. The topmost solid (black) line has $L_2=0$ and the lowest dashed (red) line has $L_2=10$. Right panel: Motion in the $(\psi,\phi)$ plane for the same choice of parameters, for which the projections of all orbits become degenerate.}
 \label{fig:thetapsiphi1}
\end{figure}

\begin{figure}[h]
\centering
 \includegraphics[width=0.75\textwidth]{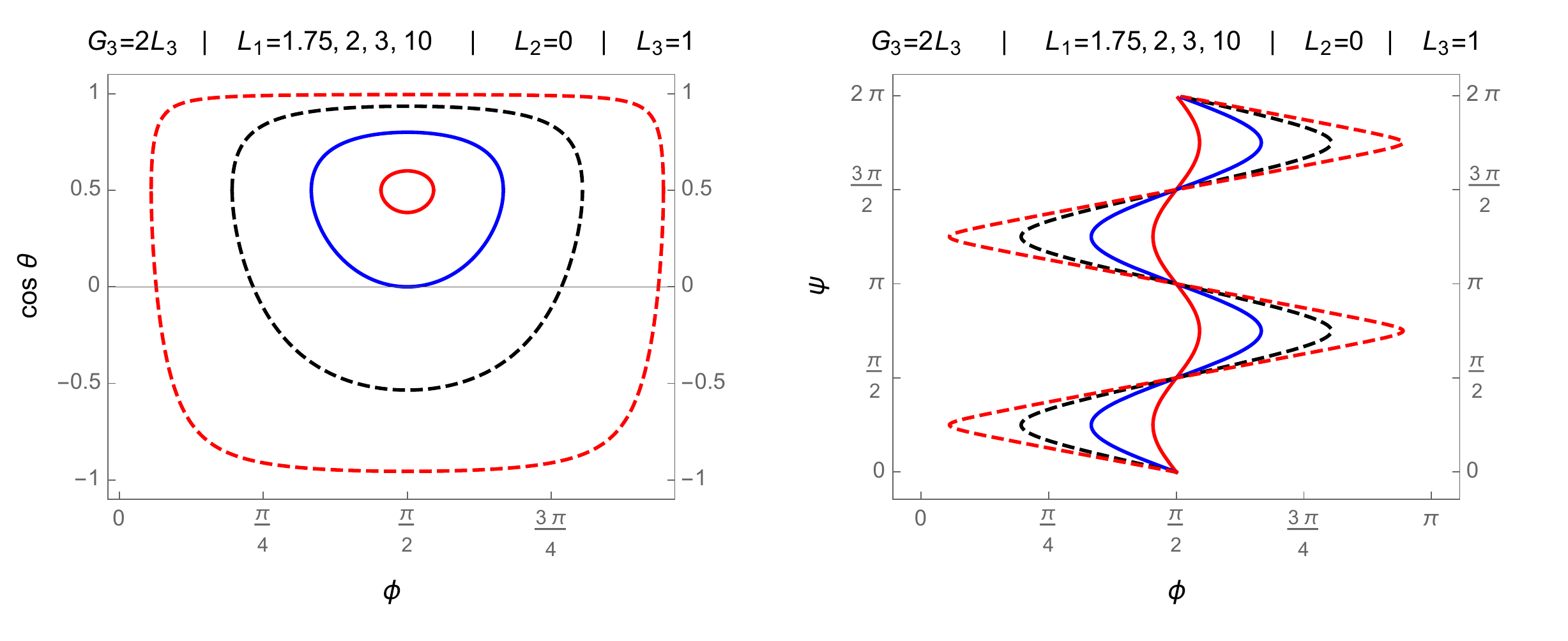}
  \vspace{-0.6cm}
 \caption{Left panel: Geodesic motion in the $(\theta,\phi)$ plane for $G_3=2L_3,\ L_3=1,\ L_2=0$ and $L_1 = 1.75, 2, 3, 10$.  The innermost solid (red) line has $L_1 = 1.75$ and the outermost dashed (red) line has $L_1 = 10$. Right panel: Motion in the $(\psi,\phi)$ plane for the same choice of parameters.}
 \label{fig:thetapsiphi2}
\end{figure}

\begin{figure}[h]
\centering
 \includegraphics[width=0.75\textwidth]{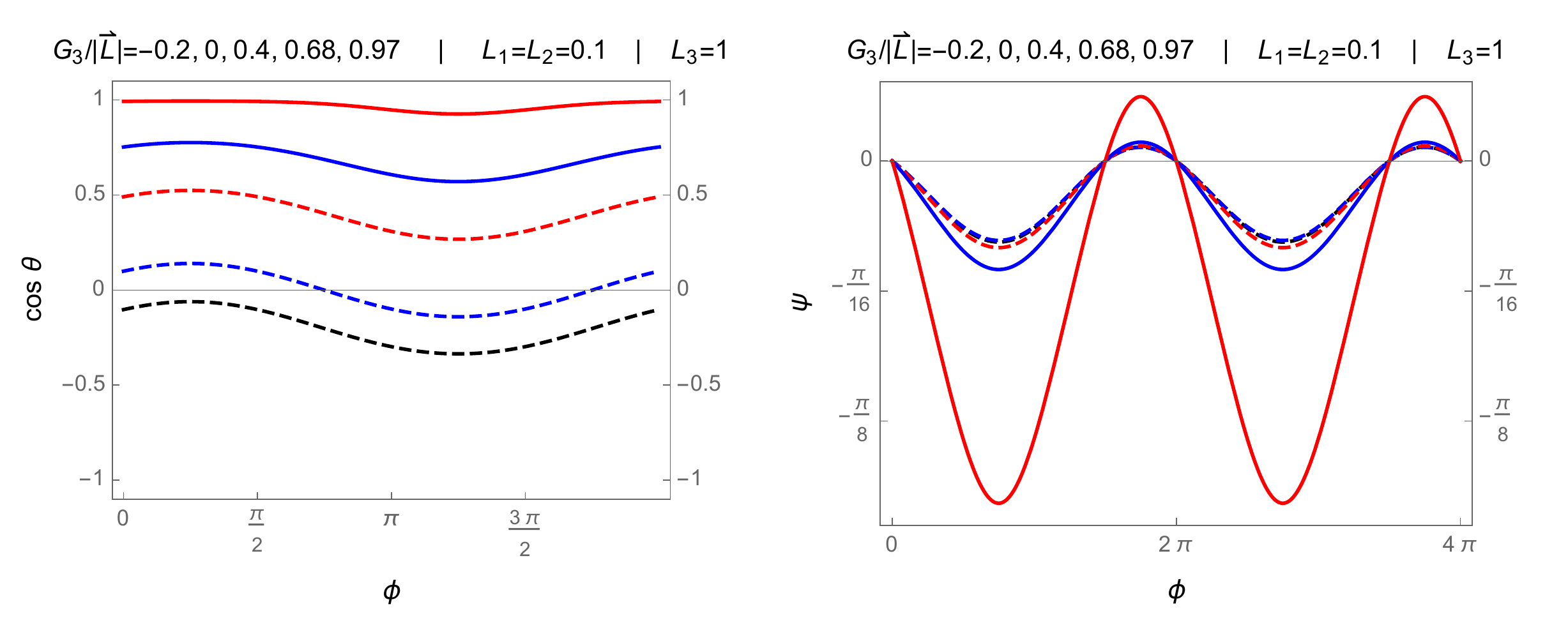}
  \vspace{-0.6cm}
 \caption{Left panel: Geodesic motion in the $(\theta,\phi)$ plane for $L_1=L_2=0.1, L_3=1$ and $G_3/|\vec L| = -0.2, 0, 0.4, 0.68, 0.97$. The topmost solid (red) line has $G_3/|\vec L| = 0.97$ and the lowest dashed (black) line has $G_3/|\vec L| = -0.2$. These orbits correspond precisely to those shown in the 3D plot of Fig.~\ref{fig:AngMotion}. Right panel: Motion in the $(\psi,\phi)$ plane for the same choice of parameters.}
 \label{fig:thetapsiphi3}
\end{figure}

\end{widetext}

Inspection of Eq.~\eqref{eq:geoU} shows that when the background is nonrotating the trajectories in the angular sector do not depend on the value of the cosmological constant $\propto \ell^{-2}$. Therefore, they are exactly the same for the asymptotically flat black holes or for their asymptotically AdS counterparts. This observation remains valid in the rotating case ($a\neq0$) but only for the projection of the trajectories in the $(\theta,\phi)$ plane.

\subsection{Innermost stable circular orbit}
\label{sec:ISCO}

Now we compute the innermost stable circular orbit (ISCO) of the nonrotating Myers--Perry--AdS black hole in $d=5$, i.e., the Schwarzschild-AdS$_5$ solution. Here we refer to ``circular'' orbits instead of ``spherical'' orbits because the spherical symmetry of the background ensures that the orbits always lie along an equatorial plane.

The radial motion for geodesics around this black hole obeys 
\beq
&& \dot r^2 + V_0(r) = 0\,, \nonumber\\
&& V_0 = \frac{\left(\ell^2 \left(r^2-2 M\right)+r^4\right) \left(L^2+\kappa  r^2\right)}{\ell^2 r^4}-E^2\,.
\label{eq:potnonrot}
\eeq
The ISCOs have a constant radial coordinate $r=r_*$ and are such that $V_0 = V_0'=V_0''=0$~\footnote{Circular orbits obviously have $V_0 = V_0'=0$, and they are stable if and only if $V_0''>0$. The innermost such orbit is marginally stable, $V_0''=0$, and corresponds to the coalescence of two branches (for sufficiently large $L$) of solutions to $V_0 = V_0'=0$, the one with largest (smallest) radial coordinate being stable (unstable). See, e.g., Refs.~\cite{Chandrasekhar:1985kt,Carroll:2004st}.}, leading to
\begin{eqnarray}
&&L^2 = 
\frac{2\kappa M r_*^2}{r_*^2-6M}\,,\\
&&E^2 = \frac{\left(L^2+\kappa  r_*^2\right) \left(\ell^2 \left(r_*^2-2 M\right)+r_*^4  \right)}{\ell^2 r_*^4}\,,\\
&&r_*^2 = 2M \left[1 + \left(\frac{M}{\mu}\right)^{\frac{1}{3}} + \left(\frac{\mu}{M}\right)^{\frac{1}{3}} \right] , \label{eq:ISCOnonrot}
\end{eqnarray}
where $\mu = M + \frac{\ell^2}{4} + \frac{\ell}{4}\sqrt{\ell^2+8 M}$.

The radius of the ISCO tends to $\sqrt{6M}$ for $\ell\rightarrow0$ and diverges like $(2M\ell)^\frac{1}{3}$ for $\ell\rightarrow\infty$. Thus, we reproduce the well-known result that there are no stable circular orbits around asymptotically flat five-dimensional (or higher) Schwarzschild black holes~\cite{Tangherlini:1963bw}.
It is worth pointing out that the existence of  stable circular orbits in the spacetimes we are considering can be attributed to the AdS asymptotics, since the radial potential~\eqref{eq:potnonrot} allows the existence of stable circular orbits even in the zero mass limit.

\section{Geodesics of the equal angular momenta Myers--Perry--AdS black hole}
\label{sec:rotating}

In what follows we shall restrict to the five-dimensional case, corresponding to $N=1$.
The spatial isometry group of the equal angular momenta black hole is given by $SU(2)\times U(1)$; one of the two $SU(2)$'s of the nonrotating case is broken down to $U(1)$. It follows that the metric thus possesses $4+1$ Killing vectors: three are associated with the $SU(2)$ subalgebra, one is associated with the $U(1)$, and another one is associated with time translation. The line element is written explicitly in the following form:
\beq
ds^2 &=& - f(r)^2 dt^2 + g(r)^2 dr^2 + \frac{r^2}{4} \left( d\theta^2 + \sin^2\theta d\phi^2 \right)\,, \nonumber\\
 && + h(r)^2 \left[ d\psi + \frac{\cos\theta}{2} d\phi - \Omega(r) dt \right]^2
\eeq
The Killing vectors that remain in the presence of rotation are $\xi_t,\xi_3,\chi_i$, defined in \eqref{eq:killing}. Again, they provide conserved quantities, which we recall here,
\be
E = -u_\mu\xi_t^\mu,\quad G_3 = u_\mu \xi_3^\mu,\quad L_i = u_\mu \chi_i^\mu,
\label{eq:EG3Li}
\ee
where $i=1,2,3$ and $u^a = (\dot t,\dot r,\dot \theta, \dot \phi,\dot \psi) $ is the velocity $5$-vector.
These first integrals allow one to solve for $\dot t,\dot \theta, \dot \phi,\dot\psi$. The solutions for $\dot\theta$ and $\dot\phi$ turn out to be the same as in the nonrotating case, Eq.~\eqref{eq:geoU}, but $\dot t$ and $\dot \psi$ acquire new terms:
\begin{eqnarray}
&&\dot t= \frac{E - G_3 \Omega(r)}{f(r)^2}\,, \label{eq:tdot}\\
&& \dot \psi = \frac{G_3}{h(r)^2}-\frac{\cos\theta \csc^2\theta}{r^2} (L_3-G_3\cos\theta) \nonumber\\
&&   \qquad \ + \frac{\Omega(r)}{f(r)^2}(E -G_3 \Omega(r))\,, \label{eq:psidot}\\
&&\dot \theta =\frac{2}{r^2} (L_1\cos\phi-L_2 \sin\phi)\,, \label{eq:thetadot}\\
&&\dot \phi = \frac{2 }{r^2}\big(L_3 - \cot\theta(L_1\sin\phi + L_2 \cos\phi)\big)\,.
\label{eq:phidot}
\end{eqnarray}
The first integrals also impose an additional constraint,
\be
G_3 = L_3 \cos\theta + \sin\theta \left( L_1 \sin\phi + L_2 \cos\phi \right)\,.
\label{eq:constraintG3}
\ee
This can be equivalently written as
\be
G_3 = \vec v \cdot \vec L\,,
\label{eq:constraint}
\ee
where $\vec L = (L_1,L_2,L_3)$ and
\be
\vec v = (\ \sin\theta \sin\phi ,\ \sin\theta \cos\phi ,\ \cos\theta \ )
\ee
is a unit vector.\footnote{The inner product in Eq.~\eqref{eq:constraint} and the statement that $\vec v$ is a unit vector refer to three-dimensional Euclidean space where the angular momentum $\vec L$ and the vector $\vec v$ live.}

\begin{figure}
 \centering
 \includegraphics[scale=.5]{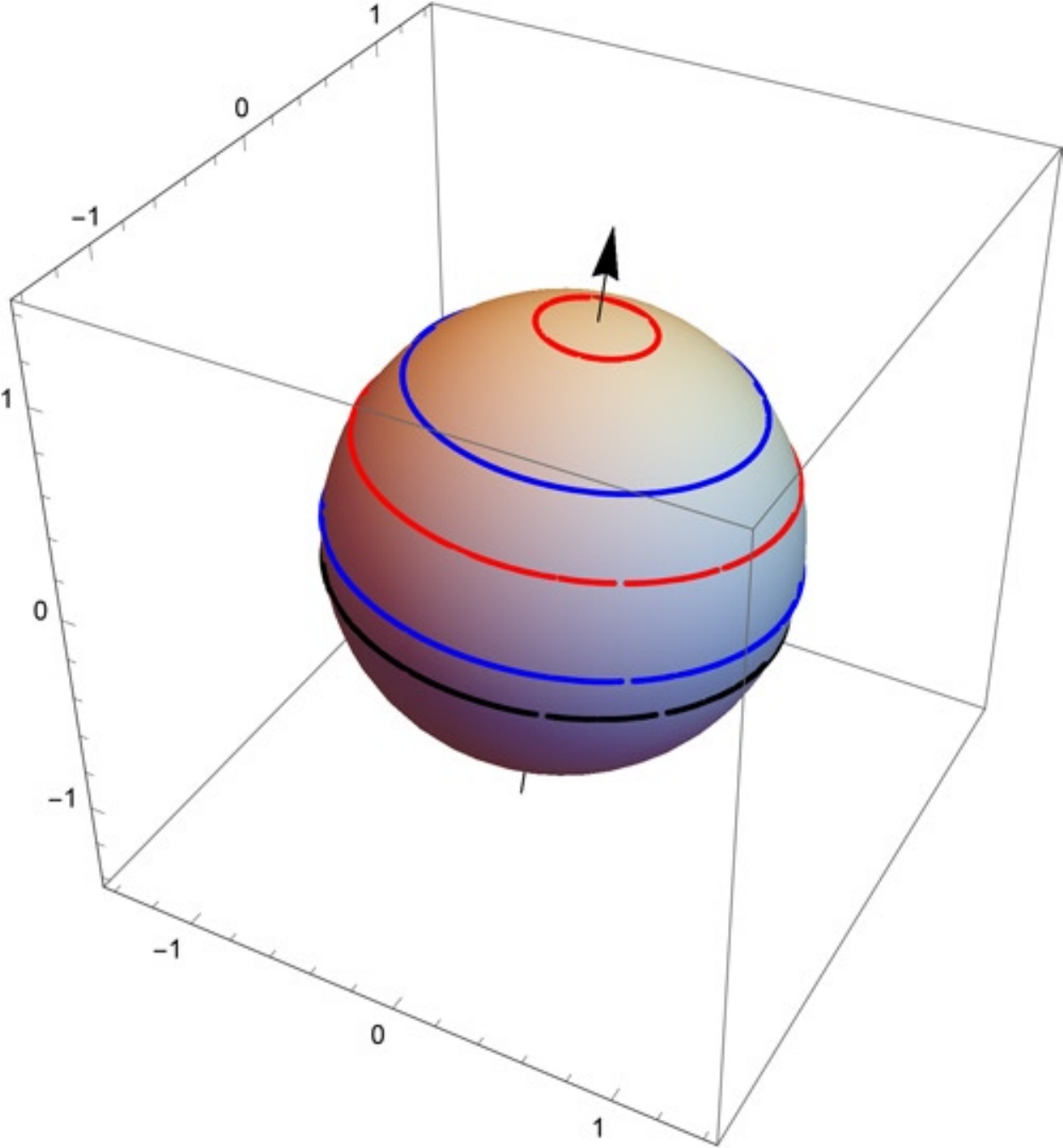}
 \caption{The motion in the $(\theta,\phi)$ sector. The motion lies in a plane intersecting the unit sphere at a latitude $\arccos(G_3/|\vec L|)$. The arrow represents the direction of $\vec L$ [which was chosen as $\vec L=(0.14, 0.14, 1.4)$ for this figure], the trajectories have $G_3/|\vec L| = -0.2, 0, 0.4, 0.68, 0.97$ from bottom to top, and the sphere has the unit radius.}
 \label{fig:AngMotion}
\end{figure}

Note that, even though expression~\eqref{eq:constraintG3} depends on angular coordinates, $G_3$ is in fact a constant of motion on geodesics, as can be straightforwardly checked by using the equations of motion~\eqref{eq:thetadot} and~\eqref{eq:phidot}. The vector $\vec v$ is a unit radial vector on $S^2$ parametrized by the colatitude $\theta$ and the longitude $\phi$. The constraint~\eqref{eq:constraint} imposes the motion on this sphere to maintain a constant angle with the angular momentum vector $\vec L$. Furthermore, the relation implies that $G_3\leq |\vec L|$. This is illustrated in Fig.~\ref{fig:AngMotion}.

As a consistency check, when $L_3=G_3\cos\theta$ and $L_1=L_2\tan\phi$, equations~(\ref{eq:tdot}-\ref{eq:phidot}) reproduce those obtained for the particular class of geodesics studied in Ref.~\cite{Rocha:2014gza}, which have $\dot\theta=\dot\phi=0$. The constraint~\eqref{eq:constraint} imposes $L_2=\sin\theta \cos\phi\, G_3$, and therefore these special geodesics are simply described by $\vec L=G_3 \vec v$.

This setup is very similar to the problem of determining geodesics in the Schwarzschild spacetime, where the metric admits three spatial Killing vectors associated to the three generators of $SO(3) \cong SU(2)$. In fact, in that case it is easy to show that one can completely eliminate the angular variable from the normalization condition on the velocity vector $u$ in terms of  $l_1^2+ l_2^2+l_3^2$, where $l_i$ are the constants of motion associated to the three spatial Killing vectors of Schwarzschild.
Here the situation is almost identical, except that the spatial symmetry group is larger, namely it is given by $SU(2)\times U(1)$. It turns out that it is also possible to explicitly eliminate the angular dependence of the problem by using 
\begin{eqnarray}
&&4 L^2  = r^4(\dot \theta^2 + \sin^2\theta\, \dot\phi^2) \nonumber\\ 
&& \qquad \quad + h(r)^4 \left(\cos\theta\, \dot\phi-2\Omega(r) \dot t+2\dot\psi\right)^2,\nonumber\\
&&4G_3^2 = h(r)^4 \left(\cos\theta\, \dot\phi-2\Omega(r)\dot t+2\dot\psi\right)^2,
\end{eqnarray}
where $L^2= L_1^2+ L_2^2 + L_3^2$.

The radial equation for the geodesics, obtained from $u_a u^a = -\kappa$, where $\kappa =0$ for null geodesics or $1$ for timelike geodesics, is conveniently expressed in terms of an effective potential $V_{eff}$,
\begin{eqnarray}
&&\dot r^2 + V_{eff}=0,\nonumber\\
&&V_{eff} = -\frac{(E-G_3 \Omega(r))^2}{g(r)^2f(r)^2}+\frac{G_3^2}{g(r)^2h(r)^2} \nonumber\\
&& \qquad \quad \ \ +\frac{L^2-G_3^2}{g(r)^2r^2}+\frac{\kappa}{g(r)^2}\,.
\label{eq:Veff}
\end{eqnarray}
This equation reduces to \eqref{eq:geoda0} when $h(r) = r$ and $\Omega(r) = 0$.
The interpretation of the constants of motion is clear: $E$ is the energy of the geodesic particle (per unit mass, in the case of massive particles) and $L^2$ is the total angular momentum squared. Moreover, $G_3$ corresponds to the angular momentum in the $U(1)$ sector, i.e., the contribution from rotations in the $\psi$ direction. This is confirmed by the determination of the geodesic equations via a Hamilton--Jacobi approach, which is presented in the Appendix~\ref{sec:HJ}. In particular, the separability constant that allows to decouple the equations of motion for $r$ and $\theta$---commonly known as Carter's constant---is shown to equal $K=L^2-G_3^2$.

Once again, particularizing to the case of geodesics that are static on the 2-sphere, i.e., have constant $\theta$ and $\phi$ coordinates, we see that $L^2=G_3^2$ (since these geodesics obey $\vec L=G_3 \vec v$) and the effective potential~\eqref{eq:Veff} reproduces the one obtained in Ref.~\cite{Rocha:2014gza}.

For the case of asymptotically flat black holes ($\ell\to\infty$) the radial potential reduces to
\beq
V_{eff}^{AF} &=& (\kappa-E^2) + \frac{L^2-2M\kappa}{r^2} + \frac{2Ma^2(L^2 -G_3^2)}{r^6} \nonumber\\
&&+ \frac{2Ma^2(\kappa-E^2)-2ML^2+4Ma E G_3}{r^4}\,,
\label{eq:AFpotential}
\eeq
which is invariant under the following scaling symmetry
\beq
&& M\rightarrow \alpha M,\quad r\rightarrow\sqrt{\alpha} r,\quad L\rightarrow \sqrt{\alpha} L,\nonumber\\
&& G_3\rightarrow\sqrt{\alpha} G_3,\quad E\rightarrow E,\quad \kappa \rightarrow \kappa.
\eeq
A direct comparison with the results of Ref.~\cite{Kagramanova:2012hw} is complicated by the fact that a different coordinate system is employed. Nevertheless, it is apparent that $r^6 V_{eff}^{AF}$ is a cubic polynomial in $r^2$ and therefore may have three positive real roots, at most. An analysis of the radial potential~\eqref{eq:AFpotential} reveals that it can have also two, one or zero positive real roots, corresponding to turning points where $\dot{r}=0$. This is in full agreement with the findings of Ref.~\cite{Kagramanova:2012hw}.

\subsection{Timelike geodesics and innermost stable circular orbits}

In this section, we study timelike geodesics of equally spinning Myers--Perry--AdS$_5$ black holes. In particular, we determine the innermost stable circular orbits in this spacetime.

In accordance with the discussion of Sec.~\ref{sec:ISCO}, the marginally stable spherical orbits (those with constant $r$ coordinate) satisfy
\be
V(r_*) = 0\,,\quad V'(r_*)=0\,,\quad V''(r_*) = 0\,,
\label{eq:is}
\ee
where $r_*$ is the radial location of these orbits. 
These orbits are degenerate. Indeed, there are three independent conserved charges ($E$, $\vec L^2$, $G_3$) plus the radial location of the orbit to solve for, while there are only three equations.
The problem can be understood in the following way: for a given black hole, there is a family of marginally stable spherical orbits parametrized by one of the charges. That is, \emph{given a value} of one of the conserved charges, there is a unique solution for the marginally stable spherical orbit. This family spans a region in the $(a,r)$ plane. The \emph{innermost} stable such orbit is the lower boundary of this region, lying outside the event horizon. 

We have seen there is an excess in the degrees of freedom of the geodesics compared to the three conditions that determine marginally stable spherical orbits. The general procedure we will follow is to solve Eq.~\eqref{eq:is} for $\vec L^2,G_3,E$ as a function of $r$. Not all values of $r$ lead to real values of the conserved charges, which is physically required. Given a set of black hole parameters, the radii $r_*$ at which $\vec L^2,G_3,E$ are all real, and simultaneously the condition
\be
G_3^2\leq \vec L^2
\ee
is satisfied, determine the region where marginally stable spherical orbits exist. The actual innermost such orbits saturate this inequality. Thus, as discussed above, those geodesics have constant $\theta$ and $\phi$ (in addition to constant $r$) and are therefore circular in this coordinate system. It is then appropriate to refer to such geodesics as innermost stable circular orbits, even though the marginally stable geodesics we determine only have constant $r$ and are generically spherical. In the nonrotating limit, the region of marginally stable spherical orbits smoothly connects with the ISCO determined by Eq.~\eqref{eq:ISCOnonrot}, and the degeneracy grows as the rotation parameter is increased, up to the extremal limit.

\begin{widetext}

 \begin{figure}[hh]
\centering
 \includegraphics[width=0.7\textwidth]{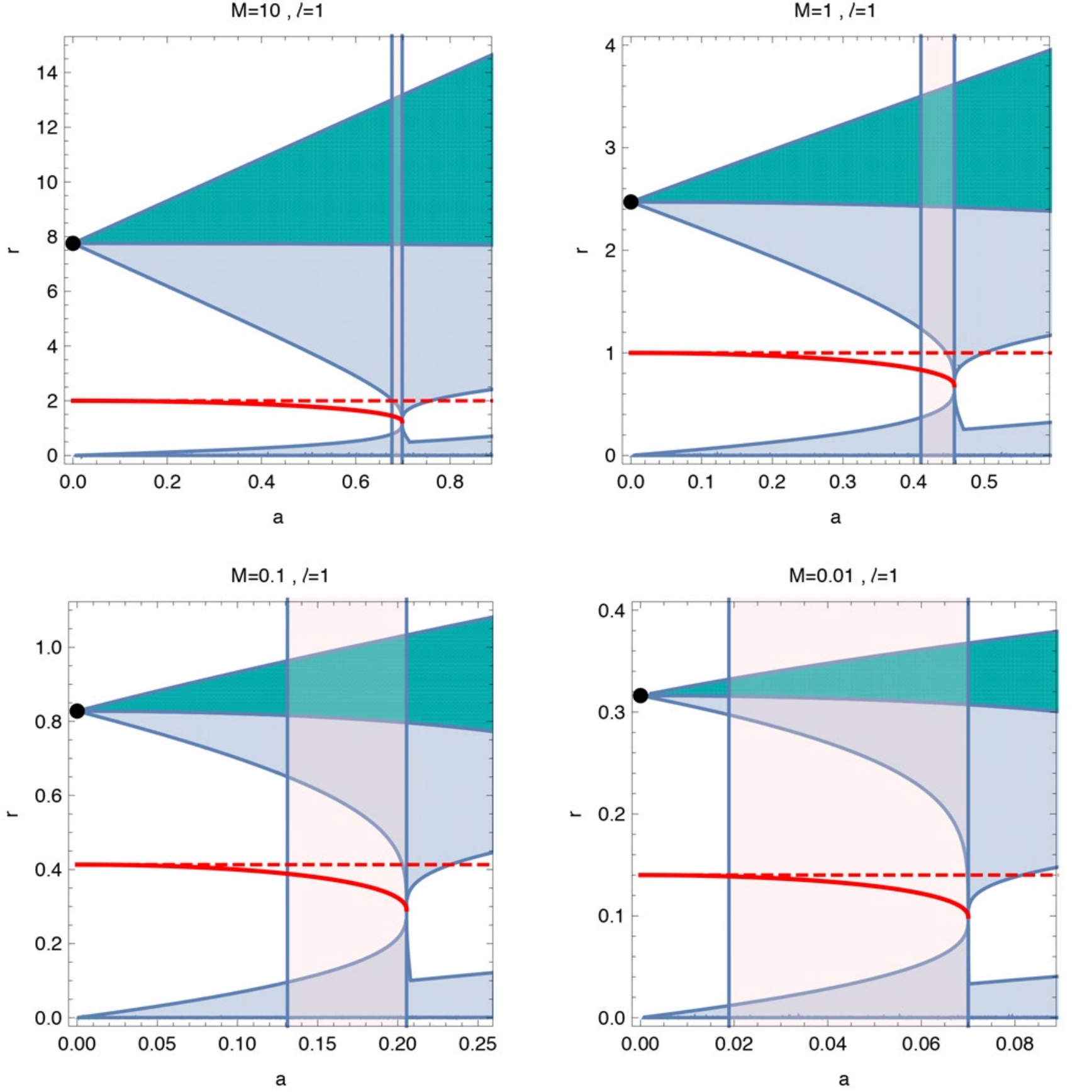}
 \caption{The innermost stable orbits for $\ell=1$ and various values of the black hole mass. The shaded gray (turquoise) area is the region where $\vec L^2,G_3,E$ are real and, in addition, $G_3^2<L^2$ and $G_3>0$ ($G_3<0$), corresponding to corotating (counter-rotating) orbits. The solid red line is the location of the event horizon, the dashed red line is the location of the ergoradius, and the shaded pink area is the region where the black hole is super-radiantly unstable. The black dot is the location of the innermost stable orbit computed directly with the nonrotating potential. The boundaries of the gray and turquoise regions---excluding the line separating the two regions---identify the limiting cases in which $G_3=|\vec L|$.}
 \label{fig:isAdS}
 \end{figure}
 
 \end{widetext}

All these features are summarized in Fig.~\ref{fig:isAdS}, where we show the region where marginally stable spherical orbits exist, the location of the horizon radius, and the ergosurface (determined by $g_{tt}=0$) as a function of the spin parameter, for several choices of the mass parameter $M$. The extremal value of the spin parameter, $a_{ext}$, is an $M$-dependent quantity and corresponds in the figure to the end point of the solid red curve identifying the black hole event horizon.

We also distinguish the orbits with respect to the sense of rotation: they are said to be corotating if $G_3>0$ and counter-rotating if $G_3<0$. This nomenclature is physically sensible since $G_3=u_\mu\xi_3^\mu$ and $\xi_3=\partial_\psi$, from Eqs.~\eqref{eq:EG3Li} and~\eqref{eq:killing}, respectively. We are therefore considering the rotation in the $\psi$ angle when we talk about co- or counter-rotation.

In addition, each panel in Fig.~\ref{fig:isAdS} includes a red shaded vertical band of which the width extends from some value of the spin parameter, $a_c$, all the way up to $a_{ext}$. This interval of rotation parameters is none other than the super-radiantly unstable regime, defined by
\be
\Omega_H \ell >1\,,
\label{eq:superradiance}
\ee
where $\Omega_H$ is the angular velocity of the black hole horizon that can be obtained by evaluating the function $\Omega(r)$ at the horizon radius.

Let us briefly discuss where condition~\eqref{eq:superradiance} comes from.
Recall that the presence of an ergoregion near the black hole event horizon leads to super-radiant scattering: an incident wave with definite frequency $\omega$ and axial quantum number $m$ is {\em amplified} as it scatters whenever $\omega<m\Omega_H$~\cite{Zeldovich:1972} (see also Ref.~\cite{Brito:2015oca}). If the black hole sits inside a confining box---like AdS spacetime---the super-radiant phenomenon might lead to an actual instability of the spacetime, since amplified waves by the black hole are reflected at the boundary of AdS~\footnote{Recall that the light-crossing time for the entire anti-de Sitter spacetime is finite.} and return to the black hole only to be amplified once again. The process then repeats itself without end and the amplitude of the field grows exponentially in time. However, it can happen that the spectrum of the field does not include modes that satisfy the super-radiant condition $\omega<m\Omega_H$, and indeed in Ref.~\cite{Hawking:1999dp} it was shown by energetic arguments that asymptotically AdS black holes satisfying the condition $\Omega_H\ell<1$ are guaranteed to be stable.
More recent studies have identified instabilities of these equal angular momenta Myers--Perry--AdS$_5$ black holes as soon as this condition is violated. Namely, Ref.~\cite{Kunduri:2006qa} found linear instabilities for a minimally coupled scalar field and~\cite{Cardoso:2013pza} found linear instabilities in the scalar gravitational sector. Hence, equally spinning black holes in the regime~\eqref{eq:superradiance} are super-radiantly unstable.

The most obvious result inferred from Fig.~\ref{fig:isAdS} is that timelike stable spherical orbits in Myers--Perry--AdS$_5$ with equal angular momenta {\em exist} outside the horizon. This is not in contradiction with the findings of Ref.~\cite{Frolov:2003en}, which showed that stable circular orbits (in the equatorial plane) do not exist in the five-dimensional Myers--Perry black hole, because the asymptotics of the spacetimes considered are different.
We have explicitly checked that, fixing $M$ and letting $\ell$ grow, the region of marginally stable spherical orbits diverges from the horizon, for sub-extremal spin parameters. Presumably, in the limit $\ell\to\infty$  this region is sent to infinity and disappears entirely.
As in the nonrotating case, the AdS confining potential is responsible for the existence of these stable spherical orbits, while the presence of a nonzero mass term ensures that such orbits do not extend all the way down to the horizon.

We remark that for black holes rotating close to extremality the ISCO submerges into the ergoregion~\footnote{This statement applies to both ``small'' or ``large'' AdS black holes, i.e., to black holes of which the horizon radius is small or large when compared to the AdS length scale $\ell$.}. For small AdS black holes this only occurs in the superradiantly unstable regime. Perhaps more interesting is the situation for large black holes, for which the ISCO can be located within the ergoregion, while the spin parameter is sufficiently low to ensure the spacetime background is stable against superradiance. 

Finally, our explorations indicate that the innermost stable circular orbit is separated from the event horizon when the black hole is subextremal, but in the extremal limit, the two surfaces appear to merge, modulo numerical resolution issues. A similar situation is known to occur in the four-dimensional Kerr spacetime~\cite{Chandrasekhar:1985kt}. Recall that extremality corresponds to the roots of the function $g(r)^{-2}$ becoming degenerate, which may then be inserted into the expression for the radial potential $V_{eff}$. However, given the high degree of the polynomials involved when $d>4$, it seems unlikely that an analytic formula can be obtained for the location of the ISCO, even in the extremal limit.

\subsection{Null geodesics}

Null orbits are characterized by $\kappa=0$. In this case, defining $\mathcal L = L/G_3$ and $\mathcal E = E/G_3$, the radial potential is given by
\begin{align}
&\frac{V_0(r)}{G_3^2} = -\mathcal E^2 \left(\frac{2 a^2 M}{r^4}+1\right) + \frac{4 a \mathcal E M}{r^4} - \frac{2 a^2 M }{\ell^2r^6}\left(\ell^2+r^2\right) \nonumber\\
&\quad \quad +\frac{\mathcal L^2}{\ell^2 r^6} \left(2 a^2 M \left(\ell^2+r^2\right)+\ell^2\left(r^4-2 M r^2\right)+r^6\right)\,.
\end{align}

The (constant-$r$) null geodesics at $r=r_c$ are such that $V_0(r_c) = 0 = V_0'(r_c)$. This leads to the simple conditions
\beq
&&\hspace{-0.9cm} \mathcal E = {\mathcal E}_\pm(M, a, r_c) \equiv \pm \frac{2 a \sqrt{M} \left(\ell^2 \left(r_c\pm\sqrt{M}\right)+r_c^3\right)}{\ell^2 \left(2 a^2 M+r_c^3 \left(r_c\pm2\sqrt{M}\right)\right)}\,, \\
&&\hspace{-0.9cm} \mathcal L^2 = {\mathcal L}_\pm^2(M, a, r_c) \nonumber\\
&&\hspace{-0.9cm} \equiv \frac{2 a^2 M \left[2 a^2 \ell^2 M+r_c^3 \left(\ell^2 \left(3 r_c\pm 4 \sqrt{M}\right)+2 r_c^3\right)\right]}{\ell^2 \left(2 a^2 M+r_c^3 \left(r_c\pm2 \sqrt{M}\right)\right)^2}\,.
\eeq

Here again, there is a family of spherical null orbits parametrized by one of the charges. We note that $\mathcal E, \mathcal L$ are real over the whole parameter space. The upper sign in the expression for $\mathcal E$ yields---for any choice of black hole parameters $(M,a)$ such that $a>0$---a solution such that $\mathcal E >0$. This is not always the case for the expression with the lower sign, for which $\mathcal E$ becomes negative for sufficiently large radius $r_c$. However, note that a solution $\mathcal E<0$ outside the ergoregion can be physically sensible (i.e., can have $E>0$) as long as $G_3$ is negative.

\begin{widetext}
 
\begin{figure}[hh]
\centering
 \includegraphics[width=0.7\textwidth]{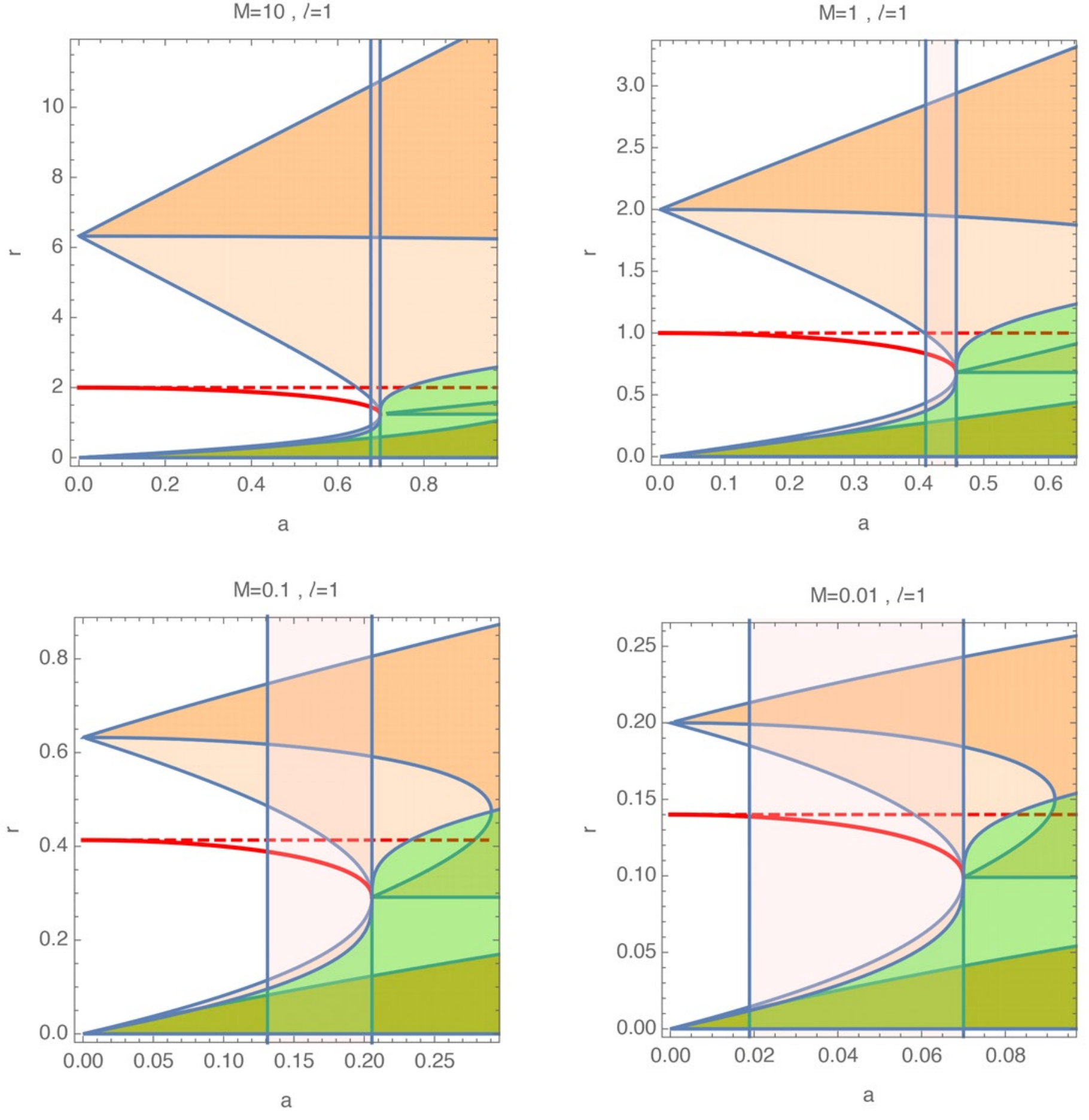}
 \caption{The null spherical orbits for $\ell=1$ and $M =10,\ 1,\ 0.1,\ 0.01$.  The orange-colored area is the region where $\mathcal L^2\geq1$. The three different shades of orange correspond to regions where ${\cal E}={\cal E}_- < 0$ (topmost region, mostly), ${\cal E}={\cal E}_- > 0$ (central vertical region, mostly), and ${\cal E}={\cal E}_+ > 0$ (bottommost region). Once again, the solid red line is the location of the event horizon, the dashed red line is the location of the ergoradius, and the vertical pink bands indicate the super-radiantly unstable backgrounds. The shaded green region in the bottom-right corner of each panel is the region where $V''>0$, i.e., where these spherical orbits are stable. In particular, there are no stable null geodesics outside the horizon: such orbits can be stable only inside the horizon or around naked singularities.}
 \label{fig:IscoNull}
\end{figure}

\end{widetext}

The criterion leading to a restriction on the location of the null orbits is ${\mathcal L}^2\geq1$, which follows from $G_3^2\leq L^2$. For solutions with $\mathcal E=\mathcal E_+>0$, this criterion is obeyed only inside the horizon (or else, for overextremal backgrounds corresponding to naked singularities). The criterion ${\mathcal L}^2\geq1$ is met for both cases $\mathcal E=\mathcal E_->0$ and $\mathcal E=\mathcal E_-<0$, corresponding to corotating ($G_3>0$) or counter-rotating ($G_3<0$) null orbits, respectively. This is illustrated in Fig.~\ref{fig:IscoNull}.

As for the timelike geodesics, the (unstable) null spherical orbits (lying outside the horizon) all collapse to a single radius in the nonrotating limit. Furthermore, at extremality the innermost null spherical orbit also appears to merge with the horizon.

Finally, let us consider the stability of these null orbits. A spherical orbit is stable if it sits at a radial location $r=r_c$ at which $V''(r_c)>0$. The region of stable spherical orbits is also shown in Fig.~\ref{fig:IscoNull}. We find that all null spherical orbits outside the black hole horizon are unstable, just as in the asymptotically flat case studied in Ref.~\cite{Frolov:2003en} for equatorial motion in a five-dimensional black hole background with arbitrary spin parameters. (This result also applies to equatorial motion around singly spinning Myers--Perry black holes in higher dimensions~\cite{Cardoso:2008bp}). Moreover, the innermost null spherical orbit (sitting outside the horizon) is always lower than the innermost timelike circular orbit, for any given choice of spin parameter $a$. We find stable null spherical orbits inside the black hole event horizon (actually, inside the Cauchy horizon), in agreement with Ref.~\cite{Diemer:2014lba}. A curious observation made possible by extending the parameter scan beyond the extremal limit is that stable null spherical orbits also exist around naked singularities, in some cases inside the ergoregion.

\section{Conclusion and discussion}
\label{sec:Conc}

In this paper, we investigated the geodesic structure of the Myers--Perry black hole in five dimensions with equal angular momenta. This black hole has a $SU(2)\times U(1)$ spatial isometry group, enhanced to $SU(2)\times SU(2) \cong SO(4)$ when the rotation vanishes. The geodesics in the static case generically admit seven (not all independent) conserved charges in the nonrotating case, of which three are associated to the generators of the first $SU(2)$, three are associated to the generators of the second $SU(2)$, and one is associated to time translation.
In the rotating case, one of the two $SU(2)$'s is broken to $U(1)$, and two conserved quantities disappear. More specifically, only the conserved quantity associated to the diagonal $U(1)$ subgroup of $SU(2)$ survives. These conserved quantities correspond to the two angular momenta vectors in the two independent rotation planes and the energy of the test particles. 
Moreover, in the static case, the two $SU(2)$'s are such that their Casimir operators are the same; i.e., the squared angular momenta of the test particle are equal.

The choice of a coordinate system well adapted to the problem under study, namely the analysis of test particles moving in this equal angular momenta spacetime, has enabled us to conclude that the angular part of the trajectories is unaffected by the cosmological constant when rotation is absent and that trajectories in the $SU(2)$ sector (parametrized by coordinates $\theta$ and $\phi$) are even independent of the spin parameter.

We have analyzed the radial motion in terms of the effective potential, which generally depends on all conserved quantities but is independent of the angular coordinates. As expected, the radial effective potential deforms continuously from the static to the rotating case.
We confirmed that there are no stable circular orbits around the five-dimensional Schwarzschild black hole, as was previously known. When extending this result to the AdS case, we found that the location of the ISCO is finite for a finite AdS radius and diverges for growing AdS radius, i.e., in the flat limit.

Stable spherical orbits continue to exist in the rotating, asymptotically AdS case. In fact, there exists a whole family of such orbits. The three independent constants of motion can be solved as a function of the radius for a marginally stable spherical orbit. Not all radii provide real values of the conserved charges, which determines---for each choice of spin parameter---a finite interval where marginally stable spherical orbits exist. The lowest radius values in this region are the actual innermost stable circular orbits and they saturate the inequality $G_3^2\leq L^2$.

We found stable spherical orbits down to the event horizon when the extremal limit is taken. Orbits that are close to the horizon can in particular be located inside the ergosphere. When considering over-rotating solutions we observed that {\em stable} null (or timelike) spherical orbits exist, possibly inside the ergoregion. The mere existence of such orbits might suggest that these nakedly singular geometries might actually be unstable. At present this issue is still unclear and certainly deserves further study.

Our study of the radial motion focused on {\em spherical} orbits, in part because they are connected to the stability of the background spacetime, at least in certain cases~\cite{Cardoso:2008bp}. Naturally, other kinds of trajectories also exist, such as unbound orbits. The type of orbit will be entirely determined by the number and location of the zeros of the radial effective potential, as in Refs.~\cite{Kagramanova:2012hw,Diemer:2014lba}, but a full exploration of this remains to be done. Nevertheless, bound orbits deviating from the spherical orbits we have studied by small oscillations are expected to produce similar qualitative results.

Finally, we have restricted to the five-dimensional case but the analysis can be extended to higher odd spacetime dimensions, although no qualitative differences are expected to appear.

\section*{Acknowledgments}

We would like to thank Thomas Basile, Vitor Cardoso and Roberto Emparan for useful discussions. T.D. acknowledges financial support from {\it Fonds National de la Recherche Scientifique} (FRS--FNRS).
J.V.R. acknowledges financial support provided by {\it Funda\c{c}\~ao para a Ci\^encia e Tecnologia} (FCT)-Portugal through Contract No.~SFRH/BPD/47332/2008, as well as under the European Union's FP7 ERC Starting Grant ``The dynamics of black holes: testing the limits of Einstein's theory,'' Grant No.~DyBHo256667.

\appendix

\section{Hamilton--Jacobi procedure
\label{sec:HJ}}

To check our results and also to be able to compare them with others in the literature, we will now derive the geodesic equations found above using a different approach: the Hamilton--Jacobi procedure. The Hamiltonian can be written as~\cite{Chandrasekhar:1985kt}
\be
{\cal H} = \frac{1}{2} g^{\mu\nu} p_{\mu} p_{\nu}\,,
\ee
where $p_{\mu}$ are the conjugate momenta. Since we are dealing with an axisymmetric and stationary spacetime---the metric equation \eqref{eq:metric} does not depend on the coordinates $t, \psi$, and $\phi$---we have the following conserved quantities:
\beq
p_t &=& g_{tt} \dot{t} + g_{t \psi} \dot{\psi} + g_{t \phi} \dot{\phi} = -E\,, \nonumber\\
p_{\psi} &=& g_{t \psi} \dot{t} + g_{\psi \psi} \dot{\psi} + g_{\psi \phi} \dot{\phi} = L_{\psi}\,, \nonumber\\
p_{\phi} &=& g_{t \phi} \dot{t} + g_{\psi \phi} \dot{\psi} + g_{\phi \phi} \dot{\phi} = L_{\phi}\,,
\label{eq:system_p}
\eeq
where the dot represents the derivative with respect to the affine parameter. We also have $p_{r} = g_{r r} \dot{r}$ and $p_{\theta} = g_{\theta \theta} \dot{\theta}$. Substituting these relations in the Hamiltonian, we get
\beq
&& \hspace{-0.3cm} \frac{h(r)^2 \left[(L_{\psi} \cot\theta-2 L_{\phi} \csc\theta)^2+4 p_{\theta}^2+\kappa  r^2\right]+L_{\psi}^2 r^2}{r^2 h(r)^2} \nonumber\\
&& + \frac{p_r ^2}{g(r)^2} - \frac{(E-L_{\psi} \Omega(r))^2}{f(r)^2} = 0 \,.
\eeq

In the above expression we have used the fact that the Hamiltonian itself is a conserved quantity determined by the normalization condition: ${\cal H} = -(1/2) \kappa$, with $\kappa =0$ for null geodesics or $1$ for timelike geodesics. To use the Hamilton--Jacobi procedure, one defines $p_r = dS_r(r)/dr$ and $p_{\theta} = dS_{\theta}(\theta)/d\theta$. The Hamilton--Jacobi equation is then written as
\beq
\frac{(L_{\psi} \cot\theta-2 L_{\phi} \csc\theta)^2+\kappa  r^2+4 S_\theta'(\theta )^2}{r^2} \nonumber\\
-\frac{(E-L_{\psi} \Omega(r))^2}{f(r)^2}+\frac{S_r'(r)^2}{g(r)^2}+\frac{L_{\psi}^2}{h(r)^2} = 0\,.
\eeq
The above equation can be written in the following way:
\beq
&&r^2 \left(\kappa -\frac{(E-L_{\psi} \Omega(r))^2}{f(r)^2}+\frac{S_r'(r)^2}{g(r)^2}+\frac{L_{\psi}^2}{h(r)^2}\right)=\nonumber\\
&&-\left[(L_{\psi} \cot\theta-2 L_{\phi} \csc\theta)^2+4 S_\theta'(\theta )^2\right]\,,
\label{eq:HJ}
\eeq
where the left-hand side is a function of $r$ only and the right-hand side is a function of $\theta$ only. Therefore, both sides must be equal to a constant, $-K$. Solving for the right-hand side, we get (choosing the positive sign)
\be
\dot{\theta} = \frac{2 \sqrt{K-(L_{\psi}  \cot\theta-2 L_{\phi}  \csc\theta)^2}}{r^2}\,.
\label{eq:thetadotApp}
\ee

The other equations for $t, \psi$ and $\phi$ are obtained by solving the system~\eqref{eq:system_p},
\beq
\dot{t} &=& \frac{E-L_\psi \Omega(r)}{f(r)^2}\,, \\
\dot{\psi} &=& \frac{L_\psi}{h(r)^2} + \frac{r^2 \Omega(r) (E-L_\psi \Omega(r))}{r^2 f(r)^2} \nonumber\\
&& +\frac{f(r)^2 \cot\theta \csc\theta (L_\psi \cos\theta-2 L_\phi)}{r^2 f(r)^2} \,, \\
\dot{\phi} &=& -\frac{2 \csc ^2 \theta (L_\psi \cos\theta-2 L_\phi)}{r^2}\,.
\eeq

The effective potential is found by using the left-hand side of Eq.~\eqref{eq:HJ}. Solving for $\dot{r}$ and then writing $V_{eff} = - \dot r^2$, we get
\be
V_{eff}(r) = \frac{\kappa -\frac{(E-L_\psi \Omega(r))^2}{f(r)^2}+\frac{L_\psi^2}{h(r)^2}+\frac{K}{r^2}}{g(r)^2}\,.
\label{eq:potentialApp}
\ee
Finally, note that we recover the relations obtained in Sec.~\ref{sec:rotating} from Eqs.~(\ref{eq:thetadotApp}--\ref{eq:potentialApp}) if the following identification between constants is made:
\be
L_{\psi} = G_3\,, \qquad
L_{\phi} = \frac{1}{2}L_3\,, \qquad
K = L^2 - G_3^2\,.
\ee


\end{document}